\PassOptionsToPackage{table,xcdraw}{xcolor}
\documentclass[11pt,a4paper]{article}
\pdfoutput=1
\usepackage{jheppub}
\usepackage{gensymb}
\DeclareSymbolFont{matha}{OML}{txmi}{m}{it}
\DeclareMathSymbol{\varv}{\mathord}{matha}{118}
\usepackage{amsmath}
\usepackage{amssymb}
\usepackage{graphicx}
\usepackage{subfigure}
\usepackage{pstricks}
\usepackage{bm}
\usepackage{pbox}
\usepackage{placeins}
\usepackage{booktabs}
\usepackage[T1]{fontenc}
\usepackage{footnote}
\usepackage{pdfpages}
\usepackage{hhline}
\usepackage{multirow}
\usepackage{multicol}
\usepackage{enumitem}
\usepackage[toc,page]{appendix}
\allowdisplaybreaks
\usepackage{comment}
\usepackage{float}
\usepackage{slashed}
\usepackage{xcolor}
\usepackage{textcomp}
\usepackage{multirow}
\usepackage[numbers]{natbib}
\usepackage{notoccite}
\usepackage[capitalise]{cleveref}
\usepackage{bm}
\usepackage{extarrows}

\usepackage{physics}
\usepackage{comment}
\usepackage{dsfont}
\usepackage{blindtext}
\usepackage{csquotes}
\usepackage[none]{hyphenat}
\DeclareUnicodeCharacter{0304}{ }
\usepackage{rotating}
\usepackage{tabularx}

\usepackage[many]{tcolorbox}
\definecolor{fg}{RGB}{34,139,34}

\renewcommand{\thetable}{\Roman{table}}

\definecolor{MyDarkBlue}{rgb}{0.1, 0.1, 0.8} 
\definecolor{MyLightBlue}{rgb}{0.22,0.51,0.9}
\definecolor{MyGreen}{rgb}{0.0, 0.5, 0.0}
\definecolor{BrickRed}{rgb}{0.8, 0.25, 0.33}

\RequirePackage{hyperref}
\hypersetup{colorlinks=true, citecolor=MyGreen,linkcolor=BrickRed, urlcolor=MyLightBlue}

\title{\bf Resolving $W$ Boson Mass Shift and CKM Unitarity Violation in Left-Right Symmetric Models with Universal Seesaw}
\author[]{K.S. Babu,}
\author[]{Ritu Dcruz}
\affiliation[]{Department of Physics, Oklahoma State University, Stillwater, OK 74078, USA}
\emailAdd{babu@okstate.edu}\emailAdd{rdcruz@okstate.edu}

\abstract{We explore the possibility of resolving simultaneously the $W$-boson mass shift reported by the CDF collaboration and the apparent deviation from unitarity in the first row of the CKM mixing matrix in a class of left-right symmetric models.  
The fermion masses are generated in these models through a universal seesaw mechanism, utilizing vector-like partners of the usual fermions.  We find a unique solution to the two anomalies where the mixing of the down quark with a vector-like quark (VLQ) resolves the CKM unitarity puzzle, and the mixing of the top quark with a VLQ partner explains the $W$ boson mass shift.  The validity of this setup is tested against the stability of the Higgs potential up to higher energies. We find upper bounds of $(5,\,4,\,4)$ TeV on the masses of the down-type VLQ, the top-partner VLQ, and a neutral scalar associated with $SU(2)_R$ gauge symmetry breaking, respectively. This class of models can solve the strong CP problem via parity symmetry without the need for an axion.
}
\begin{document}
\maketitle

\begin{sloppypar}
\section{Introduction}

Left-right symmetric models (LRSM) are well-motivated extensions of the Standard Model (SM) that provide a natural understanding of the origin of parity violation~\cite{ Pati:1974yy, Mohapatra:1974gc, Senjanovic:1975rk}.  In these models, the gauge symmetry is enhanced to $SU(3)_c \times SU(2)_L \times SU(2)_R \times U(1)_{B-L}$, with the left-handed quarks and leptons transforming as doublets under $SU(2)_L$ while the right-handed ones are doublets of $SU(2)_R$. The observed V-A structure of weak interactions is a low energy manifestation of spontaneous breaking of parity (P) symmetry which is well defined in LRSM. Apart from providing insight into the origin of parity violation, these models, owing to the presence of right-handed neutrinos to complete the gauge multiplets, also explain small neutrino masses via the seesaw mechanism, either type-I~\cite{Minkowski:1977sc,GellMann:1980vs, Mohapatra:1979ia, Yanagida:1980xy,Glashow:1979nm}, or type-II~\cite{Mohapatra:1980yp,Schechter:1980gr,Cheng:1980qt,Magg:1980ut}. Moreover, they give a physical interpretation of the hypercharge quantum numbers of fermions as a quantity arising from the $B-L$ (baryon number minus lepton number) charges and the third components of the right-handed isospin. A class of left-right symmetric models also provides a solution to the strong CP problem via parity symmetry alone, without the need to introduce the axion~\cite{Babu:1989rb}.  It is this class models that is the focus of the present paper.

The strong CP problem could very well be called the strong P problem since the large neutron electric dipole moment (EDM) induced by the QCD $\theta$-parameter is odd under parity symmetry.  In a left-right symmetric framework where parity can be defined, the QCD $\theta$-parameter would vanish owing to P.  Even after spontaneous (and soft) symmetry breaking, the quark flavor contribution to the observable $\overline{\theta}$ can be zero at the tree-level.  In the class of models studied in Ref.~\cite{Babu:1989rb}, it was shown that  $\overline{\theta}$ remains zero even after one-loop radiative corrections are included.  Small and finite $\overline{\theta}$ would arise only via two-loop diagrams, which are consistent with neutron EDM limits~\cite{Babu:1988yq,Hall:2018let}. For early work on addressing the strong CP problem via parity symmetry see~\cite{Mohapatra:1978fy,Beg:1978mt}, and for related work see~\cite{Barr:1991qx}.

The purpose of this paper is to investigate the possibility of explaining two experimental anomalies, viz., the $W$-boson mass shift and the apparent violation of unitarity in the first row of the CKM matrix, in the context of these LRSM models, while maintaining the parity solution to the strong CP problem. The masses of quarks and leptons in this class of models arise through a universal seesaw mechanism~\cite{Davidson:1987mh,Babu:1988mw}.  This is achieved by  introducing vector-like fermions (VLF)~\cite{Berezhiani:1983hm,Dimopoulos:1983rz} which are singlets of $SU(2)_L$ and $SU(2)_R$ gauge symmetries.  The SM fermions mix with these VLFs via Yukawa interactions involving a Higgs doublet $\chi_L$ of $SU(2)_L$ and its parity partner $\chi_R$, a Higgs doublet of $SU(2)_R$.   The scalar sector of the model is thus very minimal, consisting only of these two Higgs doublets. This minimality of Higgs fields plays a crucial role in solving the strong CP problem since, via gauge rotations, their vacuum expectation values (VEVs) can be both made real, resulting in vanishing contributions to $\overline{\theta}$ from the flavor sector at the tree-level.  
The light fermion masses, which are induced via a generalized seesaw mechanism, are quadratically dependent on the Yukawa couplings ($\mathcal{Y}_i$), allowing for the values of $\mathcal{Y}_i$ required to explain fermion mass hierarchy to be in the range $\mathcal{Y}_i=(10^{-3}-1)$ as opposed to $\mathcal{Y}_i=(10^{-6}-1)$ in the SM or in the standard LRSM. This class of models has received considerable attention recently in the context of flavor physics and low energy experimental signals~\cite{Craig:2020bnv}, gravitational waves~\cite{Graf:2021xku}, neutrino oscillations~\cite{Babu:2022ikf} and cosmological baryogenesis~\cite{Harigaya:2022wzt}. High-scale realizations of such models with exact parity symmetry have been developed in Ref.~\cite{Hall:2018let,Dunsky:2019api}. It is noteworthy that the neutrinos can be naturally light Dirac particles, or pseudo-Dirac particles in this context, with their masses arising from two-loop diagrams~\cite{Babu:1988yq,Babu:2022ikf}.
 
The presence of direct Yukawa couplings between VLF and SM fermions can give rise to flavor-changing neutral current~(FCNC) processes arising at the tree-level and modify the SM charged current interactions. These deviations from the SM could potentially be relevant to several experimental anomalies which have come to light in recent years, prominent among them being the muon anomalous magnetic moment  $(g-2)_\mu$ (for a recent review see~\cite{Aoyama:2020ynm}), $R_{K^{(*)}}$ (for a review and update see Ref.~\cite{Altmannshofer:2021qrr}), $R_{D^{(*)}}$ (see Ref.~\cite{Altmannshofer:2020oas}), the CDF $W$-boson mass shift \cite{CDF:2022hxs}, and the unitarity of the first row of the CKM matrix, sometimes referred to as the Cabibbo anomaly~(for reviews see Ref.~\cite{Belfatto:2019swo,Crivellin:2022ctt}). New physics implied by models such as LRSM could, in principle, resolve one or more of these anomalies.  However, with a fixed theoretical framework that solves the strong CP problem and with no room to add extra particles so that the models remain minimal, it is unsurprising that this class of LRSM models could not resolve all the anomalies.  Nevertheless, we find that the new physics contributions arising from the vector-like quarks in these models can indeed explain the $W$-boson mass shift and simultaneously explain the deviation from unitarity in the first row of the CKM matrix. (For attempts to resolve the $R_{D^{(*)}}$ anomaly in this context see Ref.~\cite{Babu:2018vrl}.)  We investigate the model parameters resulting in a concurrent solution to these two anomalies.

One of the fundamental predictions of the standard model is the unitarity of the Cabibbo-Kobayashi-Maskawa~(CKM) matrix which parametrizes the charged current weak interaction of the three generations of quarks. Each element of the CKM matrix is determined by combining experimental results with theoretical calculations that take care of relevant radiative corrections. The magnitudes of the CKM elements are well known, with enough evidence suggesting the complex nature of the matrix. The unitarity condition is a good consistency check on the otherwise overdetermined matrix. In recent years, with progress in experimental precision and a better handle on the theoretical uncertainties, the unitarity of the CKM matrix has been questioned. Namely, there is a sizeable deviation in the unitarity of the first row, as a result of the precise determination of $V_{ud}$. There is also a slightly less significant deviation in the unitarity of the first column. With a weighted average of $|V_{ud}|=0.97373 \pm 0.000 09$~\cite{Kirk:2020wdk} and the PDG-recommended average values of~\cite{ParticleDataGroup:2022pth} $|V_{us}|$ and $|V_{cd}|$, the non-unitarity appears as follows:
\begin{equation}
\begin{aligned}
    \Delta_{\text{CKM}}\equiv & 1-|V_{ud}|^2-|V_{us}|^2-|V_{ub}|^2&=&~(1.12\pm0.28)\times10^{-3}&~(\sim 3.9\,\sigma)
   \vspace{3mm} \\
  \Delta'_{\text{CKM}}\equiv & 1-|V_{ud}|^2-|V_{cd}|^2-|V_{td}|^2&=&~(3.0\pm1.8)\times 10^{-3}&~(\sim1.7\,\sigma).
\end{aligned}\label{CA}
 \end{equation}
These discrepancies are often referred to as the Cabibbo anomaly, a nod to the Cabibbo mixing of the two-generation model. The Cabibbo anomaly could be a clear indication of new physics~(NP) beyond the standard model~(BSM). As we shall see, the LRSM framework with a universal seesaw can resolve this anomaly when the down quark mixes with one of the VL-down-type quarks.

Another major challenge to the SM has appeared recently with a new high-precision measurement of the $W$-boson mass. The prediction of $W$-boson mass in SM depends solely on the mass of the $Z$ and the weak mixing angle, at the lowest order. Dependence on the gauge couplings and the masses of the top quark and the Higgs boson seep in as radiative electroweak corrections. These higher-order corrections have been computed precisely in the SM, allowing a consistency check against the measured $W$-boson mass. Recently, the CDF collaboration~\cite{CDF:2022hxs} at Fermilab has reported the most precise measurement of the $W$ boson mass so far:
\begin{equation}
    M_W^\textrm{CDF}= (80.4335 \pm 0.0094) \;\textrm{GeV}~(7\,\sigma \text{ deviation}). \label{CDF} 
\end{equation}
The CDF measurement deviates from the SM prediction~\cite{Awramik:2003rn} at $7\,\sigma$, and is also at odds with the prior PDG world average. If confirmed, this is another piece of evidence for new physics. The LRSM framework with a universal seesaw provides a unique solution to this deviation from the mixing of the top quark with one of the VL-up-type quarks.

Several NP models have been explored in the literature in resolving the CKM unitarity puzzle~(see Ref.~\cite{Crivellin:2022ctt} for a comprehensive list of references) as well as the CDF $W$-boson mass shift~(see Ref.~\cite{Dcruz:2022dao} and references therein). Among these are solutions to the two anomalies independently  with vector-like fermions~\cite{Crivellin:2020ebi,Belfatto:2021jhf,Crivellin:2021bkd,Manzari:2021prf,Manzari:2020eum,Belfatto:2019swo,Kim:2022zhj,Nagao:2022oin,Kawamura:2022uft,Balkin:2022glu}. We explore the simultaneous resolution of the two anomalies in the framework of LRSM with universal seesaw while maintaining parity symmetric solution to the strong CP problem.  This setup is more constraining compared to the general VLQ framework owing to two reasons. First, parity symmetry restricts the form of the mass matrix, as given in Eq.~(\ref{eq:massmatrix}) below, and second, possible mixing between the usual quarks and the VLQs is constrained by the quark masses. We do find a nontrivial solution with a vector-like down quark of mass below 5 TeV mixing with the down quark (to resolve the unitarity puzzle) and a top-partner VLQ with a mass below 4 TeV mixing with the top quark. In addition, a second scalar field associated with $SU(2)_R$ gauge symmetry breaking should be lighter than about 4 TeV for consistency of the model.

The remainder of the paper is organized as follows. In Sec.~\ref{sec:model} we provide a description of the model, where we recall the parity solution to the strong CP problem. Sec. \ref{sec:sec3} is devoted to discussing the two anomalies. In Sec~\ref{sec:Wmass} and Sec.~\ref{sec:CabibboAnomaly} we discuss the anomalies associated with $W$-boson mass measurement and CKM unitarity to establish the parameter space that can resolve these puzzles simultaneously. The main constraints on these parameters of the model are discussed in Sec.~\ref{sec:constraints}, with a unique and concurrent resolution to both the anomalies and the resulting predictions on the VLQ masses discussed in Sec.~\ref{sec:solution}. Finally, we conclude in Sec.~\ref{sec:conclude}.

\section{Model Description\label{sec:model}}

The particle spectrum of the LRSM with universal seesaw is composed of the usual SM fermions, right-handed neutrinos, and a set of vector-like fermionic partners for each of the light fermions denoted as $(U_a, D_a, E_a, N_a)$, where, the index $a$ is the family index. The SM fermions along with the right-handed neutrinos form left- or right-handed doublets, assigned to the gauge group $SU(3)_c \times SU(2)_L \times SU(2)_R \times U(1)_{B-L}$ as follows:
    \begin{equation}
        \begin{aligned}
            \mathcal{Q}_{L,i}\left(3,2,1,+\frac{1}{3}\right)=\begin{pmatrix}
            u_L\\d_L
            \end{pmatrix}_i, && \mathcal{Q}_{R,i}\left(3,1,2,+\frac{1}{3}\right)=\begin{pmatrix}
            u_R\\d_R
            \end{pmatrix}_i,\\
            \psi_{L,i}\left(1,2,1,-1\right)=\begin{pmatrix}
            \nu_L\\e_L
            \end{pmatrix}_i, && \psi_{R,i}\left(1,1,2,-1\right)=\begin{pmatrix}
            \nu_R\\e_R
            \end{pmatrix}_i,\\
        \end{aligned}\label{sm particles}
    \end{equation}
with $i=1 - 3$ being the family index. Here we follow the convention $ Q=T_{3L}+T_{3R}+\frac{B-L}{2}$ such that, $\frac{Y}{2}=T_{3R}+\frac{B-L}{2}$, thereby giving the hypercharge a physical meaning in terms of the $SU(2)_R$ and $U(1)_{B-L}$ quantum numbers. Vector-like fermions (VLFs) which are singlets under both $SU(2)_{L(R)}$ are introduced to generate masses for the fermions via a generalized seesaw mechanism:
    \begin{equation}
        \begin{aligned}
            U_a \left(3,1,1,+\frac{4}{3}\right),&& D_a \left(3,1,1,-\frac{2}{3}\right),&& E_a \left(1,1,1,-2\right),&& N_a \left(1,1,1,0\right).
        \end{aligned}\label{VL particles}
    \end{equation}

The Higgs sector of the model is comprised of a left-handed doublet and its parity partner, a right-handed doublet: 
    \begin{equation}
        \begin{aligned}
            \chi_L\left(1,2,1,+1\right)=\begin{pmatrix}
            \chi_L^+\\ \chi_L^0
            \end{pmatrix},&& \chi_R\left(1,1,2,+1\right)=\begin{pmatrix}
            \chi_R^+\\ \chi_R^0
        \end{pmatrix}.
        \end{aligned}\label{higgs}
    \end{equation}
The neutral component of $\chi_R$ acquires a vacuum expectation value $\langle \chi_R^0\rangle \equiv \kappa_R$ at a high scale, breaking the gauge symmetry down to that of the SM with the neutral component of $\chi_L$ acquiring a VEV $\langle \chi_L^0 \rangle \equiv \kappa_L\simeq 174$ GeV leading to the spontaneous breaking of the SM gauge symmetry. The Higgs potential of the model is given by
    \begin{equation}
        V=-(\mu_L^2\chi_L^{\dagger}\chi_L+\mu_R^2\chi_R^{\dagger}\chi_R)+\frac{\lambda_{1_L}}{2}(\chi_L^{\dagger}\chi_L)^2+\frac{\lambda_{1_R}}{2}(\chi_R^{\dagger}\chi_R)^2+\lambda_2(\chi_L^{\dagger}\chi_L)(\chi_R^{\dagger}\chi_R).\label{higgs potential}
    \end{equation}
The physical scalar spectrum $\{h,H\}$ arises from the mixing of the neutral fields $\sigma_L=Re(\chi_L^0)/\sqrt{2}$ and  $\sigma_R=Re(\chi_R^0)/\sqrt{2}$ with a mass matrix given by
    \begin{equation}
        \mathcal{M}^2_{\sigma_{L,R}}=\begin{bmatrix}
        2\lambda_{1_L}\kappa_L^2 && 2\lambda_2\kappa_L\kappa_R\\
        2\lambda_2\kappa_L\kappa_R && 2\lambda_{1_R}\kappa_R^2,
        \end{bmatrix},
    \end{equation}
resulting in the mass eigenvalues
    \begin{equation}
        \begin{aligned}
           M_h^2\simeq 2\lambda_{1_L}\left(1-\frac{\lambda_2^2}{\lambda_{1_L}\lambda_{1_R}}\right)\kappa_L^2, && M_H^2=2\lambda_{1_R}\kappa_R^2,
        \end{aligned}
    \end{equation}
where in the last step, we have assumed the hierarchy $\kappa_R \gg \kappa_L$.  Here, the field $h \simeq \sigma_L$ is identified as the SM-like Higgs boson of mass 125 GeV.

Under parity symmetry, which can be defined in LRSM owing to the enhanced gauge structure, the quark and lepton fields, as well as the Higgs fields transform as follows (with $F_{L,R}$ collectively denoting the vector-like fermions):
\begin{equation}
    \mathcal{Q}_L\leftrightarrow \mathcal{Q}_R, \,\, \psi_L\leftrightarrow \psi_R, \,\, F_L \leftrightarrow F_R, \,\, \chi_L \leftrightarrow \chi_R.
\end{equation}
The gauge boson fields also transform under parity: 
\begin{equation}
W_L^\pm \leftrightarrow W_R^\pm,
\end{equation}
so that the gauge couplings of the $SU(2)_L$  and $SU(2)_R$ factors obey the relation $g_L = g_R$ above the parity breaking scale.  In the parity symmetric limit, the quartic scalar couplings obey  $\lambda_{1_L}=\lambda_{1_R} \equiv \lambda_1$, although $\mu_L$ may be different from $\mu_R$ in Eq.~(\ref{higgs potential}) since parity symmetry may be broken softly by these dimension-two terms, without spoiling the solution to the strong CP problem. This is what we shall assume in this work. Such a soft breaking via $d=2$ terms is necessary to realize $\kappa_R \gg \kappa_L \neq 0$ in low-scale LRSM. (For the realization of high-scale LRSM with exact parity where $\mu_L^2 = \mu_R^2$, see Ref.~\cite{Hall:2018let}.)
 The conditions for the potential to be bounded from below, with $\lambda_{1_L} = \lambda_{1_R} = \lambda_1$, are
\begin{equation}
    \lambda_{1} \geq 0, \,\,\, \lambda_2\geq -\lambda_1.
\end{equation}

The charged gauge bosons are unmixed at tree-level in this framework, with their masses given by
    \begin{equation}
        \begin{aligned}
            M^2_{W^\pm_{L\,(R)}}=\frac{1}{2}g_{L\,(R)}^2\kappa_{L\,(R)}^2.
        \end{aligned}
    \end{equation}
Among the neutral gauge bosons, photon field $A_\mu$ remains massless while the two orthogonal fields $Z_L$ and  $Z_R$ mix with a mass matrix given as~\cite{Babu:2018vrl}:
    \begin{equation}
        \mathcal{M}^2_{Z_L-Z_R}=\dfrac{1}{2}\begin{pmatrix}
        (g_L^2+g_Y^2)\kappa_L^2 && g_Y^2\sqrt{\dfrac{g_L^2+g_Y^2}{g_R^2-g_Y^2}}\kappa_L^2\\
        g_Y^2\sqrt{\dfrac{g_L^2+g_Y^2}{g_R^2-g_Y^2}}\kappa_L^2 && \dfrac{g_R^4\kappa_R^2+g_Y^4\kappa_L^2}{g_R^2-g_Y^2}
        \end{pmatrix}
    \end{equation}
This gives rise to the neutral gauge bosons eigenstates $Z_{1(2)}$ with masses
    \begin{equation}
        \begin{aligned}
            M^2_{Z_1}\simeq \frac{1}{2}(g_Y^2+g_L^2)\kappa_L^2, & \quad \quad M^2_{Z_2}\simeq \frac{g_L^4 \kappa_R^2+g_Y^4 \kappa_L^2}{2(g_R^2-g_Y^2)}, 
        \end{aligned}\label{eqn:Z1Z2}
    \end{equation}
where the gauge boson $Z_1$ or, in the limit of small mixing, $Z_L$, is identified as the SM $Z$ boson of mass $91.18$ GeV. Here the hypercharge gauge coupling $g_Y$ is related to the $B-L$ gauge coupling $g_B$ via
\begin{equation}
g_Y^{-2} = g_R^{-2} + g_B^{-2}~.
\end{equation}


The Yukawa interactions of the charged fermions and the bare masses for the VLFs, are given in the flavor basis by the Lagrangian
    \begin{equation}
        \begin{aligned}
            \mathcal{L}_{\rm Yuk}&=
            \mathcal{Y}_L^u\bar{\mathcal{Q}}_L\tilde{\chi}_LU_R+\mathcal{Y}_R^u\bar{\mathcal{Q}}_R\tilde{\chi}_RU_L+M_U\bar{U}_LU_R\\&+\mathcal{Y}_L^d\bar{\mathcal{Q}}_L\chi_LD_R+\mathcal{Y}_R^d\bar{\mathcal{Q}}_R\chi_RD_L+M_D\bar{D}_LD_R\\&+\mathcal{Y}_L^e\bar{\psi}_L\chi_LE_R+\mathcal{Y}_R^e\bar{\psi}_R\chi_RE_L+M_E\bar{E}_LE_R+\mathrm{h.c.}
        \end{aligned}
        \label{eq:Yuk}
    \end{equation}
with $\tilde{\chi}_{L,R}=i\tau_2\chi^*_{L,R}$. 
Owing to parity symmetry, the Yukawa coupling matrices and the VLF mass matrices obey the relations
\begin{equation}
Y_L^{u,d,e} = Y_R^{u,d,e},~~M_{U,D,E} = M^\dagger_{U,D,E},
\end{equation}
which are crucial relations to solving the strong CP problem.  The Lagrangian given in Eq.~(\ref{eq:Yuk})  gives $6\times 6$ mass matrices for up-type quarks $(u,U)$, down-type quarks $(d,D)$ and charged leptons $(e,E)$ which can be written in the block form:
    \begin{equation}
        \mathcal{M}_{F}=\begin{pmatrix}
        0&\mathcal{Y}_{L}^f\kappa_L\\
        \mathcal{Y}^{f\,\dagger}_{R}\kappa_R & M_{F}
        \end{pmatrix}.
        \label{eq:massmatrix}
    \end{equation}

The strong CP problem is solved in these models via parity symmetry as follows.  The QCD $\theta$-parameter, $\theta_{QCD}$, being parity odd, vanishes in the model.  The physical parameter that contributes to the neutron EDM is
\begin{equation}
\overline{\theta} = \theta_{QCD} + {\rm arg}{\rm Det}({\cal M}_U {\cal M}_D)~.
\end{equation}
It is clear from the form of Eq.~(\ref{eq:massmatrix}) that when ${\cal Y}_L^f = {\cal Y}_R^f$ via parity, Det$({\cal M}_U {\cal M}_D)$ is real, and thus $\overline{\theta} = 0$ at tree-level.  Since the restriction from neutron EDM is rather severe, $\overline{\theta} \leq 10^{-10}$, it is necessary to ensure that radiative corrections do not generate a value  that exceeds this limit.  Ref.~\cite{Babu:1989rb} has shown that in this model the induced $\overline{\theta}$ vanishes at one-loop.  Very small $\overline{\theta}$ would be induced via two-loop diagrams, which have been estimated to be consistent with neutron EDM limits \cite{Babu:1989rb,Hall:2018let}. 

Quantum gravitational corrections are expected to violate all global symmetries.  Since parity symmetry falls under this category, one should worry about the quality factor of the parity solution to the strong CP problem.  The leading operator that could be induced via quantum gravity that can generate $\overline{\theta}$ is the $d=5$ operator
\begin{equation}
{\cal L}^{d=5} = \frac{1}{M_{\rm Pl}}(\overline{Q}_L Q_R) \chi_R^\dagger\, \chi_L.
\end{equation}
This operator would induce non-hermitian entries in the quark mass matrix. Taking the coefficients of these operators to be of order one and demanding that the complex contribution to the up-quark mass not generate $\overline{\theta}$ larger than $10^{-10}$ would imply that $\kappa_R \leq 10^{5}$ GeV.  This would lead to a relatively light $W_R$ boson, and the associated vector-like fermions of the model, which opens up the exciting possibility of exploring them at colliders, and possibly explaining some of the experimental anomalies.

The mass matrices of Eq.~(\ref{eq:massmatrix}) can be block diagonalized by bi-unitary transformations. The light fermion mass matrices have the form
\begin{equation}
M^f_{\rm light} \simeq - {\cal Y}_L^f M_F^{-1} {\cal Y}_R^{f \dagger} \kappa_L \kappa_R,
\end{equation}
if $|{\cal Y}_R^f \kappa_R| \ll |M_F|$ is assumed.  For a single generation, this yields the light fermion mass as $m_{f} \simeq -{\cal Y}^f_L {\cal Y}_R^f \kappa_L \kappa_R/M_F$, which is the seesaw formula now applied to charged fermions. It is also possible that there could be sizable mixing between the light fermions and the vector-like fermions.  It is such mixings that enable us to explain the $W$-boson mass shift as well as the CKM unitarity violation.  As we show in the next section, the mixing of $t$-quark with its vector-like partner can generate sufficient custodial $SU(2)$ violation in order to induce the oblique parameter $T$ and thus explain the $W$-boson mass shift.  The CKM unitarity violation would arise via mixing of the down quark with a vector-like quark.

\section{Simultaneous Solution to $W$ Boson Mass Shift  and  Cabibbo Anomaly}
\label{sec:sec3}
In this section, we briefly discuss the anomalies associated with the $W$-boson mass shift measured by the CDF collaboration and the apparent CKM unitarity violation. We also explore the parameter space required to resolve each anomaly independently in the LRSM with a universal seesaw. The flavor structure and the mixing required in each case are summarized, along with the potential constraints arising from the model framework. Finally, we arrive at a unique flavor structure that can resolve the two anomalies simultaneously within the model. It will be shown that the down quark mixing with the VL-down-type quark and the top mixing with the VL-up-type quark provides a concurrent solution.
\subsection{\texorpdfstring{$W$ Boson Mass Shift}{W Boson Mass Shift}\label{sec:Wmass}}
The CDF collaboration has recently updated the measurement of $W$-boson mass~(cf: Eq.~\eqref{CDF}) with a very high precision that has not been achieved previously~\cite{CDF:2022hxs}. This measurement is considerably at odds with the PDG world average $M_W^{PDG}=(80.379\pm0.012)$~\cite{ParticleDataGroup:2022pth} GeV, which is a combination of 
 the LEP~\cite{ALEPH:2013dgf}, Tevatron~\cite{CDF:2013dpa}~(CDF~\cite{CDF:2012gpf} and D0~\cite{D0:2012kms}) and LHCb~\cite{LHCb:2021abm} measurements, at about $3.6\,\sigma$. It is also at $\sim 7\,\sigma$ tension with the SM prediction of $M_W^\textrm{SM}= (80.357 \pm 0.004)\;\textrm{GeV}$ \cite{Awramik:2003rn}. Assuming that the CDF measurement will be confirmed by future experiments, the LRSM model can address this mass shift from the perspective of VLQ-induced corrections to the oblique parameters, $S$, $T$ and $U$ ~\cite{Peskin:1991sw}. These radiative corrections to the gauge boson propagators, for instance, in the SM appear from top and bottom quarks in the loop. This could easily be enhanced by the presence of a VL-top or -bottom quarks which have significant mixings with the usual quarks. The shift in the $W$-boson mass arising from  the oblique parameters is~\cite{Peskin:1991sw}:
 \begin{equation}
     M_W^2=M^2_{W^{SM}}+\frac{\alpha c^2}{c^2-s^2}M_Z^2\left[-\frac{1}{2}S+c^2T+\frac{c^2-s^2}{4s^2}U\right]\label{eq:Wmass}
 \end{equation}
 where, $s=\sin\theta_W$ and $c=\cos\theta_W$, $\theta_W$ being the Weinberg angle, and $\alpha$ is the fine structure constant. The oblique parameters are computed in terms of the two-point functions given as (see Appendix.~\ref{App-01} for details):
 \begin{equation}
    \begin{aligned}
        \alpha S&= 4 c^2 s^2 \left[\Pi_{ZZ}'(0)-\frac{c^2-s^2}{c s}\Pi_{\gamma Z}'(0)-\Pi_{\gamma \gamma}'(0)\right]\\
        \alpha T&= \frac{1}{c^2 M_Z^2}\left[\Pi_{WW}(0)-c^2 \Pi_{ZZ}(0)\right]\\
        \alpha U&= 4s^2\left[\Pi_{WW}'(0)-c^2\Pi_{ZZ}'(0)-2c s\Pi_{\gamma Z}'(0)-s^2\Pi_{\gamma\gamma}'(0)\right].
    \end{aligned}\label{eq:STU}
\end{equation}
Violations of custodial $SU(2)$ symmetry arise in these models via the mixing of $SU(2)_L$-doublet quark fields with singlet quarks.  The corrections to $S$ and $T$ from VLFs have been studied in great detail in~Ref.~\cite{Lavoura:1992np,Dawson:2012di,Chen:2017hak,Cao:2022mif}. We perform an independent analysis, verifying that a positive shift to $W$ boson mass would require a correction from VL-top contribution to the $T$ parameter. 

For small $t-U$ and $d-D$ mixing angles, the corrections to $S$ and $U$ are comparable and much smaller than $T$. To explain the $\sim 7 \,\sigma$ shift in $W$-boson mass (for a central value of $\Delta T=0.1726$ with $\Delta S=0=\Delta U $), we find that the mixing, say, between top and VL-top quark has to be $\sin\theta_L\gtrsim 0.1$. We shall address this quantitatively in Sec. \ref{sec:solution}, after discussing the CKM unitarity puzzle.
\subsection{CKM Unitarity Puzzle\label{sec:CabibboAnomaly}}

In recent years, with the precise determinations of $|V_{ud}|$ and $|V_{us}|$, the unitarity of the first row of the CKM is being tested. Currently, the most precise extraction of $|V_{ud}|$, aided by the progress in calculation of radiative corrections~\cite{Seng:2018yzq,Czarnecki:2019mwq,Seng:2020wjq}, comes from the super-allowed $0^+\to 0^+$ nuclear $\beta$ decays~\cite{Hardy:2014qxa,Hardy:2018zsb}. $V_{us}$ is determined from the semi-leptonic $K$ meson decays  $K\to\pi \ell \nu$~($K_{\ell 3}$, with $\ell={e,\mu}$) which involves lattice-QCD calculation of the $K\to\pi$ form factor. It can also be obtained by comparing the radiative decay rates of kaon and pion~(see Ref~\cite{Moulson:2017ive}). A significant deficit in the first row unitarity has been reported as a result of improved precision in the calculation of the radiative corrections~($\Delta_R$) to nuclear $\beta$ decay~\cite{2103.05549,Kirk:2020wdk}. The SM prediction of the CKM matrix leads to the unitarity in the first row:
\begin{equation}
    |V_{ud}|^2+|V_{us}|^2+|V_{ub}|^2= 1
\end{equation}
With the contribution from $|V_{ub}|=3.94(36)\times 10^{-3}$ being negligible, the above expression reduces to the unitarity in a two-generation model parametrized by the Cabibbo mixing angle. Using the more precise $ |V_{ud}|=0.97373(9)$~ \cite{ParticleDataGroup:2022pth,Kirk:2020wdk} and the PDG average for $|V_{us}|=0.2252(5)$ the deviation from unitarity in the first row turns out to be about 3.9$\,\sigma$~\cite{Kirk:2020wdk} as shown in Eq.~\eqref{CA}. Different approaches in determining the radiative corrections and choices of the experiments involved in estimating the CKM elements lead to inconsistency among the moduli, however, all of them show a deficit from the unitarity of about $\sim 3.3-4\,\sigma$. There is also a slightly less significant deficit in the unitarity of the first column. We address both of these in our work, although the focus is primarily on the first row non-unitarity. 

 One possible explanation of the Cabibbo anomaly stems from the idea that the CKM mixing is a sub-matrix of a more general, unitary matrix arising from the mixing of SM quarks with extra undiscovered quarks. Exploiting the mixing of SM and VL-quarks~(VLQ), we explore the scenarios inducing non-unitarity in the current model. The deviation can arise from mixing between the VLQ and light quarks in the up-sector, down-sector, or both. For illustrative purposes, let us assume that the major correction to the CKM matrix (or Cabibbo mixing matrix in case of two family mixing) arises from only one flavor of SM fermion mixing with the corresponding VLQ (referred to as VLQ-mixing, henceforth). The Cabibbo mixing can be thought of as small corrections induced by the diagonalization of the light quark matrix. The structure of the charge current interaction would heavily depend on the VLQ-mixing. We can safely ignore the case where the VLQ mixing appears from the third family as the contribution to non-unitarity would be negligible owing to the smallness of $V_{ub}$ element. VLQ-mixing arising from either the first or second generations can provide large enough correction to resolve the CKM unitarity problem. For instance, the VLQ-mixing appearing from down quark mixing with VL-down type quark can lead to a charge current structure given by
\begin{equation}
    \begin{pmatrix}
    V_{ud}~c_{L} &&V_{us}\\
    V_{cd}~c_{L}&&V_{cs}
    \end{pmatrix} \text{ or } 
    \begin{pmatrix}
    V_{ud}~c_{L} &&V_{us}~c_{L}\\
    V_{cd}&&V_{cs}
    \end{pmatrix},
\end{equation}
depending on whether the Cabibbo mixing appears from the down-sector or up-sector, respectively. Here, $c_L=\cos\theta_L$, which parametrizes the leakage of the left-handed up (down) quark into the VLQ sector, which being singlets of $SU(2)_L$ have no direct couplings to the $W^\pm$ boson. There are, in general, six different possible ways of inducing non-unitarity in the first row. A VLQ-mixing of $\sin\theta_{L}\simeq0.034$ arising from up or down quark mixing with the corresponding VLQ can resolve the  $\sim 3.9\,\sigma$ deficit in the first row unitarity, while $\sin\theta_L\simeq 0.15$ would be required if the VLQ-mixing arises from the charm or strange quark. Of these, only the VLQ-mixing arising from the first generation will survive the constraint from hadronic decays of the $Z$ boson~(see Sec.~\ref{constrain1}). 

\subsection{Constraints on model parameters\label{sec:constraints}}
\subsubsection{Light fermion masses\label{constrain1}}
In the simplified one family mixing formulation, where the $d$-quark mixes with a vector-like $D$-quark, or equivalently $s$-quark mixes with its VLQ partner, the  mass of $d$-quark (or $s$-quark) can be written as $m_i=(\kappa_L\kappa_R s_L s_R) M_{i}$, where $s_L$ and $s_R$ are the sines of left and right mixing angles among $d_i$ and $D_i$ quarks. The tangent functions ($t_{L,R}$) obey the relation $t_R=(\kappa_R/\kappa_L) t_L$. Therefore, with $t_L \simeq s_L$ we have
\begin{equation}
    s_L^2=\frac{\kappa_L}{\kappa_R} \frac{m_{d_i}}{M_{D_i}}.
\end{equation}
Since $\sin\theta_L$ should be of $\mathcal{O}(0.034)$ to resolve the Cabibbo anomaly, the products $\kappa_R M_{D_i}$ are severely restricted due to the smallness of the $d_i$ masses: $\kappa_R M_{D_i}\in [325,1.9\times10^5]$ GeV$^2$ to fit the $d$- and $s$-quark masses. This is in direct contradiction with the experimental limits on these parameters: $\kappa_R \geq 10$ TeV, corresponding to the 5 TeV lower bound on $M_{Z_R}$~\cite{ATLAS:2019erb} and $M_F>1$  TeV, from the production and decay of pair produced VLF partners~\cite{ATLAS:2018ziw,CMS:2018zkf}. Note that parity symmetry plays an important role to arrive at this conclusion.  Indeed, within the same model, without parity, the relation $t_R=(\kappa_R/\kappa_L) t_L$ will not hold, which would enable achieving large $d_L-D_L$ mixing.  In such an attempt, however, the parity solution to the strong CP problem would be lost.

The aforementioned constraints from light fermion masses can be evaded with parity symmetry intact by considering the mixing of the down quark (or the strange quark) with two VLQs.  In such a setup large $d_L-D_L$ mixing can be realized even in the limit of vanishing $d$-quark mass, as will be discussed in Sec. \ref{sec:solution}.  Small $d$-quark mass can be generated via perturbations to the mass matrix proposed there.  This scenario, where the light quark masses arise as perturbative corrections, however, is not applicable in the case of the top quark, since its mass is not so small. So, we also include the fit to the mass of the top quark in our analysis.

\subsubsection{\texorpdfstring{$Z$ Decay Width}{Z decay width}\label{constrain2}}

Another major constraint on the mixing of $d_L$ with $D_L$(or $s_L$ with its VLQ partner) arises from the modification of the $Z$ boson couplings to the fermions. The left-handed fermion interaction vertex of $Z$ is modified by the cosine of the VLQ mixing~($\cos\theta_L$), thereby altering the total decay width of the $Z$ boson:
\begin{equation}
    \mathcal{L}_Z\supset \frac{g}{\cos\theta_W}\overline{f}_L\gamma_\mu \left(T_{3L} \cos\theta_L^2-Q \sin\theta_W^2\right) f_LZ^\mu \label{eq:Zint}
\end{equation}
For instance, resolving the Cabibbo anomaly with VLQ mixing from the first generation requires $\sin\theta_L=0.034$. This results in $\Delta \Gamma_Z\simeq 1~(0.8)$ MeV from down (up)-quark mixing with VLF. On the other hand, $\sin\theta_L\simeq 0.15$ in case of the second family mixing, leading to $\Delta \Gamma_Z\simeq 20$ MeV which is excluded by the experimental measurement of $Z$ decay with an uncertainty of $\Delta \Gamma_Z\leq 2.3$ MeV at $1\,\sigma$. Therefore, the Cabibbo anomaly can be resolved only by the VLQ mixing with the first family. For a general discussion of flavor non-universality with VLQ mixing with the usual quarks, see Ref.~\cite{Belfatto:2019swo}.

\subsection{Combined Solution\label{sec:solution}}
\begin{figure}
    \centering
    \includegraphics[scale=0.35]{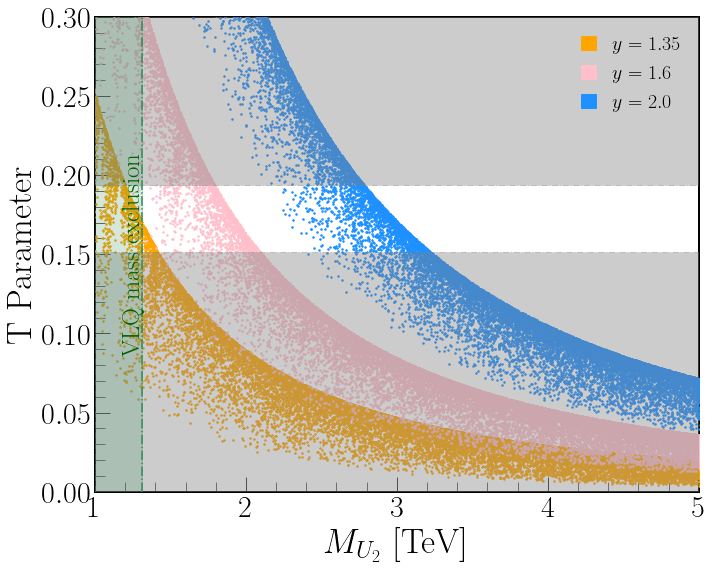}
    \caption{Oblique parameters $S,\,T,\,U$ as a function of the mass of the lightest vector-like quark ($M_{U_2}$) for different choices of the Yukawa coupling. The unshaded region shows the required range of $T$ parameter that can explain the CDF $W$-boson mass shift. The vertical dotted line shows the current experimental lower bound on VLQ mass~\cite{ATLAS:2018ziw,CMS:2018zkf}. This plot is consistent with the mass of top quark. The Yukawa coupling needs to be at least $y\simeq 1.35$ to explain the $W$-boson mass shift while evading the current experimental bound on the mass of VLQ. The corrections to  $S$ and $U$ parameters are seen to be much smaller than $T$.}
    \label{fig:stuvsmass}
\end{figure}
From the discussions so far, it is clear that the Cabibbo anomaly can only be resolved if the mixing arises from up or down quark  with a corresponding VLQ. Since the mixing angle required in this case, $\sin\theta_L\simeq 0.034$, is not large enough to accommodate the explanation for $W$ boson mass shift, which requires $\sin\theta_L\sim\mathcal{O}(0.1)$, and such large mixings are subjected to constraint from $Z$ decay width~(Sec.~\ref{constrain2}), the only remaining choice is to invoke top quark mixing with VLQ to explain the shift in the $W$-boson mass. 

As mentioned previously, a simple one-family mixing between the light and 
VL-quark suffers from a severe constraint arising from the smallness of the light fermion masses~(Sec.~\ref{constrain1}). To evade this, we assume that the down quark mixes with two vector-like quarks.  We can work in the limit of neglecting all SM fermion masses, except for the top quark.  It turns out that top-quark mixing with a single VLQ cannot induce sufficiently large $T$-parameter to explain the $W$-boson mass shift.  This is again due to parity symmetry which constrains the form of such a  $2 \times 2$ mass matrix.  When the top quark mixes with two VLQs, the contribution to the $T$-parameter can be sufficiently large, which we shall adopt.

Since there are three up-type VLQs in the model, and since the top quark mixes with two of them, there is no room for the up quark to mix significantly with any VLQs in order to explain the CKM unitarity puzzle.  We are then cornered to a unique solution that simultaneously explains the $W$-boson mass shift and the CKM unitarity puzzle: The top quark mixes significantly with two up-type VLQs, and the down quark mixes with two down-type VLQs.  The specific structures of the mass matrices in these sectors are uniquely determined and are given below:
\begin{equation}
    \begin{aligned}
    \mathcal{M}_d=\begin{pmatrix}0&&0&&y_d\kappa_L\\
     0&&0&&M_{1d}\\
    y_d\kappa_R&&M_{1d}&&M_{2d}
    \end{pmatrix}
   \text{ in $\{d~D_2~D_3\}$ basis}\\
    \mathcal{M}_t=\begin{pmatrix}0&&0&&y_u\kappa_L\\
   0 &&M_{3u}&&M_{1u}\\
    y_u\kappa_R&&M_{1u}&&M_{2u}
    \end{pmatrix}\text{ in $\{t~U_2~U_3\}$ basis}~.
    \end{aligned}
    \label{eq:mass_structure}
\end{equation}
Details of the  diagonalization of ${\cal M}_d$ are provided in Appendix.~\ref{App-02}. The up-sector diagonalization has been done numerically, with the $3 \times 3$ unitary matrix represented using the standard parametrization of the CKM matrix~(Eq.~\eqref{eq:Mu}) (where the CP violating phase is set to zero for simplicity).  Although the top-mixing also contributes to the charged current interactions, it is the down-sector mixing that resolves the Cabibbo anomaly with $\Delta_{CKM}=s_{1d}^2$ with $s_{1d} \simeq 0.034$, assuming that the CKM mixing receives a contribution from the down sector. This can be seen from the extended CKM matrix with $\left(
\overline{u}_L\,\overline{c}_L\,\overline{t}_L\,\overline{U}_{2_L}\,\overline{U}_{3_L}
\right)$ multiplied from the left and $\left(d_L\,s_L\,b_L\,D_{2_L}\,D_{3_L}\right)^T$ from the right~(see Appendix~\ref{App-02} for details):
\begin{equation}
  \begin{pmatrix}
  
\textcolor{MyGreen}{c_{{1d}} V_{ud}} && \textcolor{MyGreen}{c_{{1d}} V_{us}} && \textcolor{MyGreen}{c_{{1d}} V_{ub}} && -c_{{2d}}
   s_{{1d}} && -s_{{1d}} s_{{2d}}  \vspace{1mm}\\
\textcolor{MyGreen}{ V_{cd}}&& V_{cs} && V_{cb} && 0 && 0 \vspace{1mm}\\
\textcolor{MyGreen}{ u_{11} V_{td}} && u_{11} V_{ts} && u_{11} V_{tb} && 0 && 0 \vspace{1mm}\\
 u_{21} V_{td} && u_{21} V_{ts} && u_{21} V_{tb} && 0 &&
   0 \vspace{1mm}\\
 u_{31} V_{td} && u_{31} V_{ts} && u_{31} V_{tb} && 0 && 0
   \\
  \end{pmatrix}\label{eq:WLint}  
\end{equation}
From Eq.~\eqref{eq:WLint} one can also see that a combination of $u_{11}=\hat{s}_{12}^u$ when $\hat{s}_{13}^u\ll1$~(Eq.~\eqref{eq:Mu}) and $s_{1d}$ can explain the non-unitarity of the CKM matrix in the first column, with the main correction appearing from the down-sector owing to the smallness of $V_{td}$. The $Z$ interaction with left-handed light fermions gets modified owing to the mixing of $d$ with $D$ and $t$ with $U$. New interactions involving $\overline{t} U_i Z$ and $\overline{d} D_iZ$ arise in the model proportional to these VLQ mixings. 
 The remaining left-handed quark vertices with the $Z$ boson, and all the right-handed quark vertices with $Z$, however, remain unchanged. These new interactions by themselves do not cause any tree-level flavor changing processes.
 
Both up- and down-sectors contribute to the corrections to the oblique parameters, but the effect from down-sector is negligible compared to that of top-mixing, since the mixing of only $s_{1d}\simeq 0.034$ is required to resolve the Cabibbo anomaly. The expressions for $S$, $T$ and $U$ arising from $t-U$ mixing has been verified against~Ref.~\cite{Dawson:2012di,Chen:2017hak}. The exact expressions in our framework are given in Appendix~\ref{App-01}~(\cref{eq:T,eq:S,eq:U}). We find that the corrections to $S$ and $U$ are negligible compared to the $T$ parameter.

\begin{figure}
    \centering
    \includegraphics[scale=0.7]{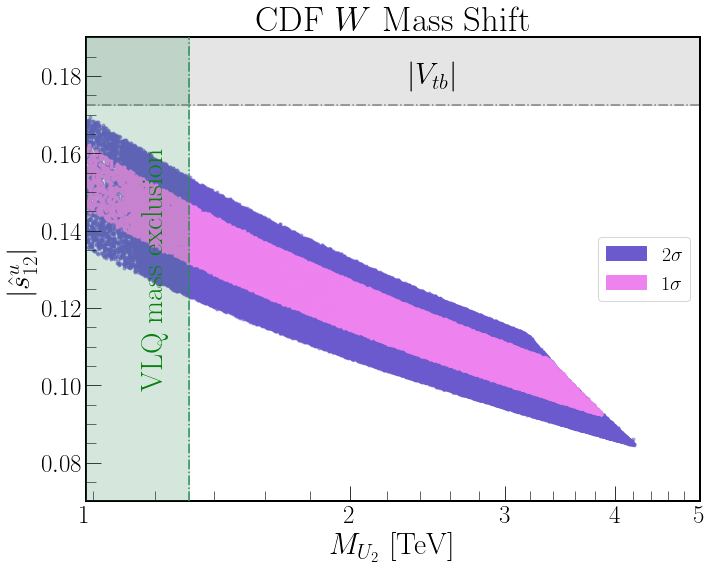}
    \caption{1- and 2-$\sigma$ regions required to explain CDF-$W$ mass and Cabibbo anomaly plotted as a function of the sine of the mixing angle $\hat{s}_{12}^u$~(Eq.~\eqref{eq:Mu}), consistent with the stability of Higgs potential. The green shaded region is the exclusion on VLQ mass~\cite{ATLAS:2018ziw,CMS:2018zkf} whereas the gray shaded region corresponds to the limit on mixing angle from the determination of $|V_{tb}|$~\cite{ParticleDataGroup:2022pth}  element of the CKM matrix.}
    \label{fig:Wmass}
\end{figure}

The behavior of the $T$ parameter as a function of the lightest VLQ mass~($M_{U_2}$) is shown in Fig.~\ref{fig:stuvsmass}, for  different choices of the Yukawa coupling $y$, imposing the fit for top quark mass. It becomes evident that the Yukawa coupling $y_u$ has to be at least 1.35 to explain the positive shift in $W$-boson mass while evading the LHC constraint on VLQ mass. The fit to the CDF $W$-boson mass shift using the full expressions of oblique parameters~(\cref{eq:T,eq:S,eq:U}) is shown in Fig.~\ref{fig:Wmass}. The region shown in the plot is also consistent with $1 \text{ and }2~\,\sigma$ allowed region from resolving the CKM unitarity puzzle from down-sector mixing, although this alone is not enough to explain the shift in $W$-boson mass and CKM non-unitarity simultaneously. In Fig.~\ref{fig:Wmass} we have also indicated the excluded region from $|V_{tb}|$ measurement as a shaded region on the top. In resolving the Cabibbo anomaly, it is essential to understand that the new elements of the CKM matrix are identified with those determined experimentally, i.e., $|c_{1d}V_{us}|\equiv |V_{us}^{PDG}|$, etc. These re-definitions, however, do not impose any constraints on these essentially free parameters. 

With only the down quark and the top quark mixing with VLQs significantly, there would not be any large FCNC mediated by the $Z$ boson or the Higgs boson.  However, once all the CKM mixing angles are induced, there could be such effects.  To control these, we note that $V_{us}$ can arise from the up-sector, while $V_{cb}$ arises from the down-sector.  These have no bearing on the light-heavy mixing.  The smaller $V_{ub}$ may arise from either sector, involving admixture of the $(u-c)$ sector in the mass matrix ${\cal M}_t$ of Eq. (\ref{eq:mass_structure}), which is not expected to lead to large FCNC effects, owing to the smallness of $V_{ub}$.

\subsection{Stability of the Higgs Potential}

The resolution to the $W$-boson mass shift, along with the necessary fit to the top quark mass requires Yukawa coupling $y_u$ to be larger than about 1.35.  Although such large Yukawa couplings are not excluded by the perturbative unitarity bounds~\cite{Chanowitz:1978uj,Marciano:1989ns,Whisnant:1994fh,Allwicher:2021rtd}, the Higgs potential is at risk of becoming unstable as the quartic coupling $\lambda_{1_L}$~(Eq.~(\ref{higgs potential})) could turn negative when extrapolated to higher energies. Partial wave unitarity would require $y_u^2< 8\pi/3$ \cite{Marciano:1989ns}, obtained from the condition Re($a_0) \leq 1/2$, for the elastic fermion-antifermion scattering in the color singlet channel, at the center of mass energies much above the fermion mass. While this condition was derived for a chiral fermion of the SM such as the top quark, it would equally apply to the scattering of vector-like top quark at energies well above its bare mass. It should be noted that the heavier VLQ with order one Yukawa coupling obtains its mass primarily via the Yukawa coupling, rather than through its bare mass.

The stability of the Higgs potential is especially of concern since the heavier VLQ  with order one Yukawa coupling has a mass of order 20 TeV in the model. This requires that $\lambda_{1_L}$  should remain positive up to this energy scale. Here, we show that the renormalization group flow of the quartic couplings in the momentum range $(4-20)$ TeV would keep $\lambda_{1_L}$ positive, which would guarantee the stability of the Higgs potential up to this energy scale. 

Suppose that the scalar field $\sigma_R$ associated with the $SU(2)_R$ symmetry breaking has a mass of order 20 TeV.  In this case, only the quartic coupling $\lambda_{1_L}$ is active below 20 TeV, which by virtue of the Yukawa coupling $y_u$ would turn negative in going from 4 TeV to this scale.  On the other hand, if $\sigma_R$ has a mass comparable to the lighter VLQ, of order 4 TeV, in the momentum range $(4 - 20)$ TeV both $\lambda_{1_L}$ and $\lambda_{1_R}$ are active.  $\lambda_{1_L}$ would decrease in going to higher energies driven by the Yukawa coupling of $\sigma_L$, while $\lambda_{1_R}$ will not feel the effect of the Yukawa coupling.  This is because the Yukawa coupling of  $\sigma_R$ involves heavier fields.  This is the scenario that is consistent with our framework.  Naturally, this scheme would predict that the $\sigma_R$ field should have a mass of the same order as the $U_2$ field, which is of order 4 TeV or lower.

To test the validity of our scenario, we have computed the RG evolution of the quartic coupling and shown that the coupling $\lambda_{1_L}\geq0$ at least up to 20 TeV. We have presented the one-loop renormalization group equations for the dimensionless parameters of the model in Appendix~\ref{sec:RGE}. These are generalizations of equations given in Ref.~\cite{Koide:2000ks,Mohapatra:2014qva} that can be adopted to the regime below 20 TeV, where parity symmetry is not exact.  From the RG evolution of the Yukawa couplings, given in Eq. (\ref{eq:RGE1}), we find that for $y_{u,d} \simeq 1.4$, there is no significant running in $y_{u,d}$ in the momentum range $(4-20)$ TeV.  Thus it is a good approximation to take $y_{u,d}$ as constants in this regime.  With this approximation, we can  analytically compute the evolution of the quartic scalar couplings.  

At the momentum scale of order $M_{U_3}$, where parity symmetry is exact, we have $\lambda_{1_L}=\lambda_{1_R}=\lambda_1$. At lower energies $\lambda_{1_L}$ will be larger than $\lambda_{1_R}$, owing to the Yukawa contributions to their evolution.  Note that $\sigma_R$ couples to the heavier VLQ, while $\sigma_L$ couples to the lighter VLQ in our fits. Therefore below the parity restoration scale, $\lambda_{1_R}$ does not feel the effects of its large Yukawa couplings.
The difference $\delta_\lambda=\lambda_{1_L}-\lambda_{1_R}$, defined at the lower scale, is obtained by solving the relevant renormalization group equations, Eqs. \eqref{eq:RGE2} and \eqref{eq:RGE2a}, applicable to the momentum range $(4-20)$ TeV.  The solution can be written down approximately as
\begin{equation}
    \delta_\lambda=\frac{3}{16\pi^2}\left(y_u^4 \ln{\left[\frac{M_{U_3}}{M_{U_2}}\right]}+y_d^4 \ln{\left[\frac{M_{U_3}}{M_{D_2}}\right]}\right).
\end{equation}
Only the Yukawa couplings cause a significant split in the $\lambda_{1_L}$ and $\lambda_{1_R}$, since the gauge couplings$g_L$ and $g_R$ run identically below 20 TeV. 
We minimize the Higgs potential at a momentum scale of order 4 TeV, which roughly corresponds to the $\sigma_R$ scalar mass. At this scale $\lambda_{1_R}$ is taken to be a free parameter.  
$\lambda_{1_L}$ obtained at the low momentum scale using SM RG evolution~\cite{Arason:1991ic} of the Higgs quartic coupling is used to compute the SM Higgs mass $m_h$ at that scale.  To ensure that the Yukawa couplings do not cross the perturbative unitarity bound of $|y|\leq \sqrt{8\pi/3}$~\cite{Marciano:1989ns} upon RG evolution~\cite{Arason:1991ic}, the coupling was chosen to $|y_u|\lesssim2.25$, for $|y_d|<1$ for the entire momentum range. A few benchmark points illustrating that $\lambda_{1_L}$ remains positive are given in Table.~\ref{tab:my_label}.

\begin{table}[]
\resizebox{\textwidth}{!}{%
\begin{tabular}{|c|c|c|c|c|c|c|c|c|c|c|c|c|c|c|c|}
\hline
 & $y_u$ & $y_d$ & \begin{tabular}[c]{@{}c@{}}\{$M_{1u}, M_{2u}, M_{3u}$\} \\ \{$M_{1d}, M_{2d}$\}\end{tabular} & $M_{D_2}$ & $M_{D_3}$ & $M_{U_2}$ & $M_{U_3}$ & $|s_{1d}|$ & $|s_{2d}|$ & $|\hat{s}^u_{12}|$ & $\lambda_{1_R}$ & $\lambda_2$ & $\lambda_{1_L}$ & $m_H$ & $T$ \\ \hline
BM-1 & 2.19 & 0.28 & \begin{tabular}[c]{@{}c@{}}\{2.9,3.0,1.3\} \\ \{1.5,0.2\}\end{tabular} & 1.5 & 3.2 & 3.2 & 22.4 & 0.033 & 0.034 & 0.11 & 0.01 & 0.09 & 0.17 & 1.5 & 0.19 \\ \hline
BM-2 & 1.97 & 0.19 & \begin{tabular}[c]{@{}c@{}}\{2.8, 2.7,1.4\} \\ \{1.4,0.3\}\end{tabular} & 1.3 & 2.4 & 3.0 & 20.0 & 0.026 & 0.093 & 0.099 & 0.008 & 0.056 & 0.17 & 1.3 & 0.15 \\ \hline
BM-3 & 2.06 & 0.64 & \begin{tabular}[c]{@{}c@{}}\{2.6, 3.6,1.2\} \\ \{4.6,3.8\}\end{tabular} & 4.0 & 9.1 & 2.9 & 21.0 & 0.024 & 0.279 & 0.11 & 0.04 & 0.15 & 0.17 & 2.9 & 0.19 \\ \hline
BM-4 & 1.92 & 0.56 & \begin{tabular}[c]{@{}c@{}}\{2.3, 2.7,1.2\} \\ \{3.8,3.8\}\end{tabular} & 3.2 & 8.0 & 2.5 & 19.6 & 0.025 & 0.284 & 0.12 & 0.031 & 0.11 & 0.17 & 2.5 & 0.19 \\ \hline
BM-5 & 2.00 & 0.24 & \begin{tabular}[c]{@{}c@{}}\{2.8, 3.0, 1.3\} \\ \{1.2,0.02\}\end{tabular} & 1.2 & 2.7 & 3.0 & 20.4 & 0.035 & 0.005 & 0.10 & 0.007 & 0.052 & 0.19 & 1.2 & 0.16 \\ \hline
\end{tabular}%
}
\caption{Benchmark points showing that $\lambda_1$ remain positive up to the momentum scale of the heaviest VLQ mass~$M_{U_3}$, for Yukawa couplings $y_u$ and $y_d$ which are chosen at the low momentum scale equivalent to the lightest of the VLQ masses. All the masses are quoted in TeV. $M_{U_{2(3)}}$ and $M_{D_{2(3)}}$ are the light~(heavy) VLQ masses in the up- and down-sectors, respectively. The heavy Higgs mass, $m_H$ is chosen to be the lowest VLQ mass to obtain $\lambda_1$ and $\lambda_2$. The value of $T$ parameter is given for each benchmark point.}
    \label{tab:my_label}
\end{table}

On applying all these constraints, we find that the largest allowed mass of up-type VLQ is $M_{U_2}\lesssim 3.8~(4.2)$ TeV, whereas the down-type mass can be $M_{D_2}\lesssim 4.2~(5.2)$ TeV, to resolve the anomalies within $1\,(2)\,\sigma$. In getting these limits, we have allowed $|y_d| = 2.25$, its perturbative unitarity limit. However, the stability of the Higgs potential is not easy to guarantee when both $|y_u|$ and $|y_d|$ are of order 1.5 or greater. If $|y_d| < 1$ is imposed, while allowing for larger $|y_u|$,  the mass of $D_2$ would be $M_{D_2}\lesssim 1.8~(2.2)$ TeV, to resolve the Cabibbo anomaly within $1\,(2)\,\sigma$. These limits are the more conservative ones, consistent with the Higgs potential stability.

\section{Conclusion \label{sec:conclude}}
In light of the recent CDF measurement of the $W$-boson mass reporting a possible $7\,\sigma$ shift with the SM prediction, and the improvements in  determination of the CKM matrix elements resulting in an apparent violation of unitarity in the first-row as large as $\sim 3.9\,\sigma$, we have investigated new physics beyond the SM. In this paper, we focus on a particular class of left-right symmetric models wherein the fermion masses are generated by a universal seesaw mechanism aided by the presence of heavy vector-like fermionic partners having Yukawa couplings to the light SM fermions. We explore the UV-complete model to provide a successful explanation of the CKM unitarity puzzle and the $W$-boson mass shift in a well-constrained parameter space. We have insisted that parity symmetry is broken only by dimension-two soft terms in the scalar potential, which would provide a solution to the strong CP problem without the need for an axion. We find the simplest, unique flavor structure that can explain the two anomalies where one of the SM quarks mixes with two VLQs. The CKM unitarity puzzle can be explained by the mixing of either or both of the first-generation quarks with VLQs while the $W$-boson mass shift requires top-quark mixing with VLQs. For a concurrent explanation, invoking the top-quark mixing eliminates the possibility of up-quark mixing with two VLQs simultaneously. This leads to a unique solution where the up-sector (top-quark mixing) and down-sector (down-quark mixing) with two VLQs each, resolve the $W$ mass shift and the CKM unitarity puzzle, respectively. Since this framework requires $\mathcal{O}(1)$ Yukawa coupling, we have also analyzed the stability of the Higgs potential. Ensuring that the potential remains stable till at least up to the mass of the second VLQ, this model predicts an upper bound on the VL up-type quark mass of $\sim 4.2 $ TeV and the VL down-type mass of $\sim 5.2$ TeV.

\appendix
\section{Mass Matrix Diagonalization}
\label{App-02}
The mass structures in Eq.~\eqref{eq:mass_structure} are diagonalized using a bi-unitary transformation of the form $U_L \mathcal{M}U_R^\dagger$. Here, $U_{L(R)}$ are parameterized by $\theta_{L(R)_{1,2}}$. Since we are interested only in the left-handed mixing angles, $U_L=U$ with $\theta_{L_{1,2}}= \theta_{1,2}$. The down-sector left-handed fields are diagonalized by $U_d\mathcal{M}_d\mathcal{M}_d^T U_d^T$, assuming all the couplings to be real. 
\begin{equation}
    \begin{aligned}
    \mathcal{M}_d\Rightarrow \text{Diag}(0,\,M_{D_2},\,M_{D_3}).
    \end{aligned}\label{eq:Md}
\end{equation}
\begin{equation}
    U_{d}=\begin{pmatrix}
    1&0&0\\
    0&c_{{2d}}&-s_{{2d}}\\
    0&s_{{2d}}&c_{{2d}}
    \end{pmatrix}\begin{pmatrix}
    c_{{1d}}&s_{{1d}}&0\\
     -s_{{1d}}&c_{{1d}}&0\\
    0&0&1
   
    \end{pmatrix}.
\end{equation}
$c(s)$ stands for $\cos(\sin)$ where, $\theta_{(1,2)d}$ are the left-mixing angles. 
The mixing angles are
\begin{equation}
    \begin{aligned}
    \theta_{1d}=\arctan(\frac{-y_d\kappa_L }{M_{1d}}),&&\theta_{2d}=\frac{1}{2}\arctan\left(\frac{2M_{2d}\sqrt{M_{1d}^2+y_d^2\kappa_L^2}}{M_{2d}^2+(\kappa_R^2-\kappa_L^2)y_d^2}\right),\\
      \end{aligned}\label{eq:mixing}
\end{equation}
and the heavy VLQ mass eigenvalues are
\begin{equation}
    \begin{aligned}
    M_{D_{3,2}}=&\frac{1}{2}\left(2M_{1d}^2+M_{2d}^2+(\kappa_L^2+\kappa_R^2)y_d^2\pm\sqrt{4M_{2d}^2(M_{1d}^2+\kappa_L^2y_d^2)+\left(M_{2d}^2+(\kappa_R^2-\kappa_L^2)y_d^2\right)^2}\right).
      \end{aligned}\label{eq:VLQmass}
\end{equation}

For the up-sector, we can  represent a general unitary matrix under the standard CKM-like parametrization (with the CP phase set to zero for simplicity) as
\begin{equation}
\begin{aligned}
     U_{u}&=\begin{pmatrix}
    u_{11}&u_{12}&u_{13}\\
    u_{21}&u_{22}&u_{23}\\
    u_{31}&u_{32}&u_{33}
    \end{pmatrix}\\
    &=\left(\begin{array}{ccc}
\hat{c}_{12} \hat{c}_{13} & \hat{s}_{12} \hat{c}_{13} & \hat{s}_{13} \\
-\hat{s}_{12} \hat{c}_{23}-\hat{c}_{12} \hat{s}_{23} \hat{s}_{13}  & \hat{c}_{12} \hat{c}_{23}-\hat{s}_{12} \hat{s}_{23} \hat{s}_{13} & \hat{s}_{23} \hat{c}_{13} \\
\hat{s}_{12} \hat{s}_{23}-\hat{c}_{12} \hat{c}_{23} \hat{s}_{13} & -\hat{c}_{12} \hat{s}_{23}-\hat{s}_{12} \hat{c}_{23} \hat{s}_{13}  & \hat{c}_{23} \hat{c}_{13}
\end{array}\right)
\end{aligned}
   \end{equation}
such that
\begin{equation}
    \mathcal{M}_t\Rightarrow \text{Diag}(m_t,\,M_{U_2},\,M_{U_3}),\label{eq:Mu}
\end{equation}

It should be noted that the mass matrices in \cref{eq:mass_structure} need to be modified by small corrections to fit the light fermion masses. The exact mass matrix diagonalization can proceed through a bi-unitary transformation with the unitary matrices $\mathcal{U}_{L(R)}$ parameterized by mixing angles $\rho_{L(R)}\ll 1$. For instance, after the diagonalization of the large couplings in the down sector, the mass matrix can take the following structure:
\begin{equation}
    \mathcal{M}'_d=\begin{pmatrix}
    x && \epsilon\kappa_L\\
    \epsilon'\kappa_R && X
    \end{pmatrix},
    \end{equation}
 where $x,\epsilon,\epsilon' \text{ and} X$ are functions of the perturbative couplings and the large mixing angles $\theta_{{L\,(R)}_{1d\,(2d)}}$. $\epsilon' =\epsilon^\dagger$ under $\{c_{L_{1d}}\to c_{R_{1d}}\,c_{L_{2d}}\to c_{R_{2d}}\,s_{L_{2d}}\to s_{R_{2d}}, \kappa_L\to\kappa_R \}$. This matrix structure can be block-diagonalized as follows:

   \begin{equation}
        \mathcal{U} _L\mathcal{M}'_d \mathcal{U}_R^\dagger=\begin{pmatrix}
        \mathds{1}-\frac{1}{2}\rho_L\rho^\dagger_L&&\rho_L\\
        -\rho^\dagger_L&&\mathds{1}-\frac{1}{2}\rho^\dagger_L\rho_L
        \end{pmatrix}\begin{pmatrix}
    x && \epsilon\kappa_L\\
    \epsilon'\kappa_R && X
    \end{pmatrix}
   \begin{pmatrix}
        \mathds{1}-\frac{1}{2}\rho_R\rho^\dagger_R&&-\rho_R\\
        \rho^\dagger_R&&\mathds{1}-\frac{1}{2}\rho^\dagger_R\rho_R
        \end{pmatrix},
    \end{equation}
    giving rise to the SM fermion mass matrix of the form
 \begin{equation}
     \hat{m}=x-\kappa_L\kappa_R \epsilon M^{-1}\epsilon'\,^\dagger,
 \end{equation}
 where $M=\text{Diag}(M_1,M_2,M_3)$ are the heavy VLQ masses. 
 \begin{equation}
     \rho_L=\kappa_L \epsilon M^{-1},\,\, \qquad\qquad\rho_R=\kappa_R\,\epsilon'\,^\dagger M^{-1},
 \end{equation}

\section{Oblique Parameters}
\label{App-01}
The expressions for oblique parameters in Eq.~\eqref{eq:STU} are obtained from~\cite{Peskin:1991sw}
\begin{equation}
    \begin{aligned}
        \alpha S&\equiv 4 e^2 \left[\Pi_{33}'(0)-\Pi_{3Q}'(0)\right]\\
        \alpha T&\equiv \frac{e^2}{s^2c^2 M_Z^2}\left[\Pi_{11}(0)-\Pi_{33}(0)\right]\\
        \alpha U&\equiv 4e^2\left[\Pi_{11}'(0)-\Pi_{33}'(0)\right]
    \end{aligned}
\end{equation}
and converting these to the gauge boson basis using~\cite{Kundu:1996ah}:
\begin{equation}
    \begin{aligned}
        \Pi_{\gamma\gamma}&=e^2\Pi_{QQ},\,\,\, \Pi_{WW}=\frac{e^2}{s^2}\Pi_{11},\\
        \Pi_{\gamma Z}&=\frac{e^2}{sc}(\Pi_{3Q}-s^2\Pi_{QQ}),\\
        \Pi_{ZZ}&=\frac{e^2}{c^2 s^2}(\Pi_{33}-2s^2\Pi_{3Q}+s^4\Pi_{QQ})
    \end{aligned}
\end{equation}
Here, $s=\sin\theta_W$ and $c=\cos\theta_W$, $\theta_W$ being the Weinberg angle, and $\alpha$ is the electromagnetic coupling strength. In deriving Eq.~\eqref{eq:STU} we have used $\Pi_{\gamma \gamma}(0)=0=\Pi_{\gamma Z}(0)$.
Other variations of these expressions can be found in~\cite{Holdom:1990xp,Bhattacharyya:1991yw,Ma:1992uc,Lavoura:1992np,Kundu:1996ah,Grimus:2008nb} among others. The difference between the expressions which are written in terms of slopes and derivatives are negligible under the approximation that $\Pi_{ij}$ are linear functions of $q^2$ near $q^2=0$, i.e.,
\begin{equation}
  \Pi_{ij}(q^2)= \Pi_{ij}(0)+q^2\Pi'_{ij}(0) 
\end{equation}
See Ref.~\cite{Kundu:1996ah} for a detailed discussion on the oblique parameters. Some expressions may be in terms of 
\begin{equation}
    \Pi_{3Y}=2(\Pi_{3Q}-\Pi_{33}).
\end{equation}
Wherever the definitions of $U$ differ in terms of $M_W$ and $M_Z$, care must be taken to use the relevant expression for $W$ mass shift. 
The complete expressions for the oblique parameters in our framework are (using the short-hand $\sin\theta_{if}=s_{if}$ and $\cos\theta_{if}=c_{if}$):

\begin{align}
   T&= \frac{N_c}{16\pi s^2 M_W^2}\Bigg[\left(u_{11}^4-1\right) m_t^2+\left(u_{21}^4 M_{U_2}^2+u_{31}^4M_{U_3}^2
   \right)\nonumber\\
    &+2\left\{u_{11}^4-1+m_t^2 \left(\frac{1-u_{11}^2}{m_t^2-m_b^2}-\frac{u_{11}^2u_{21}^2}{ M_{U_2}^2-m_t^2}-\frac{ u_{11}^2u_{31}^2}{
   M_{U_3}^2-m_t^2}\right)\right\}m_t^2 \ln{\left[\frac{m_t^2}{M_{U_3}^2}\right]} \nonumber\\
   &+ 2u_{21}^2\left\{u_{21}^2+M_{U_2}^2 \left(- \frac{
    1}{M_{U_2}^2-m_b^2}+\frac{u_{11}^2 
   }{ M_{U_2}^2-m_t^2}-\frac{u_{31}^2}{ M_{U_3}^2-M_{U_2}^2}\right)\right\}M_{U_2}^2\ln{\left[\frac{M_{U_2}^2}{M_{U_3}^2}\right]} \nonumber\\
       &+2 
   \left\{\frac{u_{21}^2+u_{31}^2}{m_t^2-m_b^2}-\frac{u_{21}^2}{M_{U_2}^2-m_b^2}-\frac{u_{31}^2}{M_{U_3}^2-m_b^2}\right\}m_b^4\ln{\left[\frac{m_b^2}{M_{U_3}^2}\right]}\Bigg],\label{eq:T}
\end{align}

\begin{align}
    S&=\frac{N_c}{18\pi}\Bigg[5\left(-u_{11}^2+u_{11}^4-u_{21}^2u_{31}^2\right)\nonumber\\
    &-\left\{1+2u_{11}^2-3u_{11}^4 -6 u_{11}^2 m_t^4\left(u_{21}^2\frac{ 3
  M_{U_2}^2-m_t^2}{\left(M_{U_2}^2-m_t^2\right)^3}+u_{31}^2\frac{  
  3 M_{U_3}^2-m_t^2}{\left(M_{U_3}^2-m_t^2\right)^3}\right)\right\} \ln{\left[\frac{m_t^2}{M_{U_3}^2}\right]}\nonumber\\
  &+u_{21}^2\left\{ 3 u_{21}^2-4+6
  M_{U_2}^4 \left( u_{31}^2 \frac{3 M_{U_3}^2-M_{U_2}^2}{\left(M_{U_3}^2-M_{U_2}^2\right)^3}+u_{11}^2\frac{   M_{U_2}^2-3 m_t^2}{\left(M_{U_2}^2-m_t^2\right)^3}\right)
\right\}\ln{\left[\frac{M_{U_2}^2}{M_{U_3}^2}\right]}\nonumber\\
&+12\left\{u_{11}^2u_{21}^2 \frac{  m_t^2M_{U_2}^2}{\left(M_{U_2}^2-m_t^2\right)^2}+u_{11}^2u_{31}^2\frac{  m_t^2M_{U_3}^2}{\left(M_{U_3}^2-m_t^2\right)^2}+ u_{21}^2u_{31}^2\frac{ M_{U_2}^2M_{U_3}^2}{\left(M_{U_3}^2-M_{U_2}^2\right)^2}\right\}\Bigg],\label{eq:S}
\end{align}

\begin{align}
  U&=\frac{N_c}{18\pi}\Bigg[ \frac{7}{2}\left(1-u_{11}^4\right)-\frac{3}{2}\left(u_{21}^4+u_{31}^4\right)-2\left(u_{21}^2+u_{31}^2-u_{21}^2u_{31}^2\right)\nonumber\\
    &+12\left\{m_b^2
   \left(\left(u_{11}^2-1\right)\frac{m_t^2}{\left(m_t^2-m_b^2\right)^2}+u_{21}^2\frac{
   M_{U_2}^2}{\left(M_{U_2}^2-m_b^2\right)^2}+u_{31}^2\frac{M_{U_3}^2
   }{\left(M_{U_3}^2-m_b^2\right)^2}\right)\right.\nonumber\\
   &\left.+m_t^2 u_{11}^2\left(u_{21}^2\frac{
   M_{U_2}^2}{\left(M_{U_2}^2-m_t^2\right)^2}-u_{31}^2\frac{M_{U_3}^2 }{\left(M_{U_3}^2-m_t^2\right)^2}\right)-u_{21}^2u_{31}^2
   \frac{M_{U_2}^2 M_{U_3}^2 }{\left(M_{U_3}^2-M_{U_2}^2\right)^2}\right\}\nonumber\\
   &+6\left\{\left(u_{11}^2-1\right)\frac{3
   m_t^2-m_b^2}{\left(m_t^2-m_b^2\right)^3 }+u_{21}^2\frac{ 3 M_{U_2}^2-m_b^2
   }{\left(M_{U_2}^2-m_b^2\right)^3}+u_{31}^2\frac{3M_{U_3}^2-m_b^2 }{
   \left(M_{U_3}^2-m_b^2\right)^3  }\right\} m_b^4\ln{\left[\frac{m_b^2}{M_{U_3}^2}\right]}\nonumber\\
   &-6\left\{\left(u_{11}^2-1\right)\frac{m_t^2-3 m_b^2}{
   \left(m_t^2-m_b^2\right)^3 }+u_{11}^2u_{21}^2 \frac{3 M_{U_2}^2-m_t^2}{
   \left(M_{U_2}^2-m_t^2\right)^3}+u_{11}^2u_{31}^2\frac{3 M_{U_3}^2 -m_t^2}{
   \left(M_{U_3}^2-m_t^2\right)^3 }\right\} m_t^4\ln{\left[\frac{m_t^2}{M_{U_3}^2}\right]}\nonumber\\
  &+6u_{21}^2\left\{\frac{ M_{U_2}^2-3 m_b^2}{ \left(M_{U_2}^2-m_b^2\right)^3 }-u_{11}^2\frac{M_{U_2}^2- 3 m_t^2}{\left(M_{U_2}^2-m_t^2\right)^3}-u_{31}^2 \frac{3 M_{U_3}^2-M_{U_2}^2}{ \left(M_{U_3}^2-M_{U_2}^2\right)^3 }\right\} M_{U_2}^4\ln{\left[\frac{M_{U_2}^2}{M_{U_3}^2}\right]}\nonumber\\
  &+3\left(1-u_{11}^4\right)\ln{\left[\frac{m_t^2}{M_{U_3}^2}\right]}-3u_{21}^2u_{21}^2\ln{\left[\frac{M_{U_2}^2}{M_{U_3}^2}\right]}\Bigg].\label{eq:U}
\end{align}
The contributions from down-sector mixing can be obtained under the transformation
\begin{equation}
    \begin{pmatrix}
u_{11}&u_{12}&u_{13}\\ u_{21}&u_{22}&u_{23}\\u_{31}&u_{32}&u_{33}
\end{pmatrix}\to\begin{pmatrix}
c_{1d}&&s_{1d}&&0\\-c_{2d}s_{1d}&&c_{1d}c_{2d}&&-s_{2d}\\-s_{1d}s_{2d}&&c_{1d}s_{2d}&&c_{2d}
\end{pmatrix}
\end{equation}
with $\{M_{U_2}\to M_{D_2},\,M_{U_3}\to M_{D_3},\,m_t\to m_d,\, \text{and } m_b \to m_u\}$

\section{One-loop Renormalization Group Equations}
\label{sec:RGE}
Here we present the full set of one-loop RGE for the dimensionless parameters of the LRSM with a universal seesaw. These generalize the equations given in Ref. \cite{Koide:2000ks,Mohapatra:2014qva}. 

The gauge couplings $g_3,\,g_L,\,g_R,$ and $g_B$ evolve with momentum according to the renormalization group equations (with $t = {\rm ln}\mu$)
\begin{equation}
16 \pi^2 \frac{d g_i}{dt} = b_i g_i^3
\end{equation}
where $b_i = (-3,\,-\frac{19}{6},\,-\frac{19}{6},\,\frac{41}{2})$ for $i=(3,\,L,\,R,\,B)$. The Yukawa coupling matrices of Eq.~\eqref{eq:Yuk} evolve according to the equations
\begin{eqnarray}
16\pi^2 \dv {{\cal Y}_A^f}{t} = \left(T_A^f - G_A^f + H_A^f\right)\,{\cal Y}_A^f, ~~~~~~A=L,R, ~~~f=u,d,e.
\end{eqnarray}
Here $T_A^f$ arise from fermion loops, $G_A^f$ from gauge boson loops, and $H_A^f$ from Higgs boson loops.  The functions $T_A^f$ and $G_A^f$ are proportional to the unit matrix, which $H_A^f$ is flavor-dependent.  The explicit forms of these functions are given by~\cite{Koide:2000ks}
\begin{eqnarray}
T_A^u &=& T_A^d = T_A^e = 3\left[{\rm Tr}({\cal Y}^u_A {\cal Y}^{u \dagger}_A) + {\rm Tr}({\cal Y}^d_A {\cal Y}^{d \dagger}_A)\right] + {\rm Tr}({\cal Y}^e_A {\cal Y}^{e \dagger}_A) \nonumber \\
G_A^u &=& \frac{17}{8}g_B^2+ \frac{9}{4}g_{2A}^2 + 8 g_3^2 \nonumber \\
G_A^d &=& \frac{5}{8}g_B^2+ \frac{9}{4}g_{2A}^2 + 8 g_3^2 \nonumber\\
G_A^e &=& \frac{45}{8}g_B^2+ \frac{9}{4}g_{2A}^2 \nonumber\\
H_A^u &=& -H_A^d = \frac{3}{2}({\cal Y}_A^u {\cal Y}_A^{u\dagger} - {\cal Y}_A^d {\cal Y}_A^{d\dagger}) \nonumber \\
H_A^e &=& \frac{3}{2}{\cal Y}_A^e {\cal Y}_A^{e \dagger}~.
\label{eq:RGE1}
\end{eqnarray}

The quartic scalar couplings of Eq.~\eqref{higgs potential} evolve with momentum according to the equations
\begin{eqnarray}
16 \pi^2 \frac{\lambda_{1_L}}{dt} &=& 12 \lambda_{1_L}^2 + 4 \lambda_2^2 + \frac{9}{4}(g_L^4 + g_L^2 g_B^2 + \frac{3}{4}g_B^4) - \lambda_{1_L} ( 9 g_L^2 + \frac{9}{2}g_B^2) \nonumber \\
&+& 4 \lambda_{1_L} \{3 {\rm Tr}({\cal Y}_L^u {\cal Y}_L^{u \dagger}) + 3 {\rm Tr} ({\cal Y}_L^d {\cal Y}_L^{d \dagger}) + {\rm Tr} ({\cal Y}_L^e {\cal Y}_L^{e \dagger})  \} \nonumber \\
&-& 4 \left\{3\left({\rm Tr}({\cal Y}_L^u {\cal Y}_L^{u \dagger})\right)^2 + 3 \left({\rm Tr} ({\cal Y}_L^d {\cal Y}_L^{d \dagger})\right)^2 + \left({\rm Tr}({\cal Y}_L^e {\cal Y}_L^{e \dagger})\right)^2\right\} 
\label{eq:RGE2}
\end{eqnarray}
\begin{eqnarray}
16 \pi^2 \frac{\lambda_{1_R}}{dt} &=& 12 \lambda_{1_R}^2 + 4 \lambda_2^2 + \frac{9}{4}(g_R^4 + g_R^2 g_B^2 + \frac{3}{4}g_B^4) - \lambda_{1_R} ( 9 g_R^2 + \frac{9}{2}g_B^2) \nonumber \\
&+& 4 \lambda_{1_R} \{3 {\rm Tr}({\cal Y}_R^u {\cal Y}_R^{u \dagger}) + 3 {\rm Tr} ({\cal Y}_R^d {\cal Y}_R^{d \dagger}) + {\rm Tr} ({\cal Y}_R^e {\cal Y}_R^{e \dagger})  \} \nonumber \\
&-& 4 \left\{3\left({\rm Tr}({\cal Y}_R^u {\cal Y}_R^{u \dagger})\right)^2 + 3 \left({\rm Tr} ({\cal Y}_R^d {\cal Y}_R^{d \dagger})\right)^2 + \left({\rm Tr}({\cal Y}_R^e {\cal Y}_R^{e \dagger})\right)^2\right\} ~.
\label{eq:RGE2a}
\end{eqnarray}

\end{sloppypar}

\section*{Acknowledgements}
This work is supported by the U.S. Department of Energy  under grant number DE-SC0016013. Some computing for this project was performed at the High-Performance Computing Center at Oklahoma State University, supported in part through the National Science Foundation grant OAC-1531128.
\bibliographystyle{utphys}

\bibliography{ref}
\end{document}